\documentclass[a4paper,12pt]{article}
\usepackage{epsfig}
\usepackage{amssymb}
\usepackage{amsmath}

\begin{document}

%
%
\begin{titlepage}
\vspace{3 ex}
\begin{center}
{
\LARGE \bf \rule{0mm}{7mm}{\boldmath NEW PHYSICS IN CP-VIOLATING OBSERVABLES FOR BEAUTY}\\
}

\vspace{4ex}

{\large J.Bernab\'eu}
\vspace{1 ex}

{\em Department of Theoretical Physics  
and IFIC,\\ University of Valencia and CSIC}

\vspace{2 ex}

\end{center}

\vspace{2 ex}
%
%
\begin{abstract}

After the present establishment of CP-Violation in $B_{d}$-physics, consistency tests of unitarity in the Standard Model and the search of new phenomena are compulsory. I illustrate the way to look for T-violation, without contamination of absorptive parts, in correlated decays in B-factories. $B_{s}$-mixing and penguin-mediated $B_{s}$-decays are of prime importance in hadronic machines to look for new physics.

\end{abstract}
\end{titlepage}

%

\newpage
\pagestyle{plain} 

\section{Introduction}\label{section1}
As exemplified by the presentations of this 2002 edition of the
''Beauty'' Conference, CP violation is currently the focus of a great deal
of attention. The results of the B-factories \cite{cita01} measure $\sin (2\beta)$,
where $\beta $ is the CP-phase between the top and charm sides of the (bd)
unitarity triangle \cite{cita02}. To within the experimental sensitivity, the CKM
mechanism \cite{cita03} of the standard theory is verified. In this description, all
the CP-violating observables depend on a unique phase in the quark mixing
matrix. Any inconsistency between two independent determinations of the
CP-phase is an indication for new physics. Two main reasons for the search
of new physics in CP-violating observables are:

i) \ The dynamic generation of the baryonic asymmetry in the Universe
requires CP-violation, and its magnitude in the standard model looks
insufficient.

ii) \ Essentially all extensions of the standard model introduce new sectors
with additional sources of CP-phases.

The examples that I will consider here to search for new physics are based
on the following attitude: the decays dominated by standard model diagrams
at tree level with W-exchange allow the extraction of $\left| V_{ub}\right| $%
, $\left| V_{us}\right| $ and $\left| V_{cb}\right| $. On the contrary, new
physics will be apparent in processes which, for the standard model, are
described by loop diagrams, like $B_{d}-\overline{B}_{d}$ and $B_{s}-%
\overline{B}_{s}$ mixing or penguin amplitudes. In so doing, I will cover
cases which are appropriate for B-factories, as well as others which need
hadronic machines.

In Section \ref{section2} I discuss coherent correlated decays of $B_{d}$'s in order to
build (besides CP-odd asymmetries) T-odd and CPT-odd observables. Section \ref{section3}
is devoted to temporal asymmetries which need absorptive parts and are a
signal of a non-vanishing $\Delta \Gamma $ in $B_{d}$ decays. In Section \ref{section4} I
discuss $B_{s}$-mixing. The decays $B_{s}\rightarrow J/\psi \phi $ and $%
B\rightarrow \phi K_{S}$ are presented in Section \ref{section5} as ways to search for
new physics. Some conclusions are drawn in Section \ref{section6}.
\bigskip

\section{Genuine Asymmetries from Entangled States}\label{section2}

In a B-factory operating at the $\Upsilon (4S)$ peak, correlated
pairs of neutral B-mesons are produced. This permits the performance of
either a flavour tag or a CP tag. To $O(\lambda ^{3})$, where $\lambda $ is
the Wolfenstein parameter \cite{cita04} in the quark mixing matrix, the determination
of the single $B$-state is possible and unambiguous \cite{cita05}. Any final
configuration (X,Y), where X,Y are decay channels which are either flavour
or CP conserving, corresponds to a single particle mesonic transition. The
intensity for the final configuration
\begin{equation}
I(X,Y,\Delta t)\equiv \frac{1}{2}\int_{\Delta t}^{\infty }dt^{\prime }\left|
(X,Y)\right|^{2}
\end{equation}
is thus proportional to the time dependent probability for the meson
transition.

\begin{figure}[h]
\begin{center}
\epsfig{figure=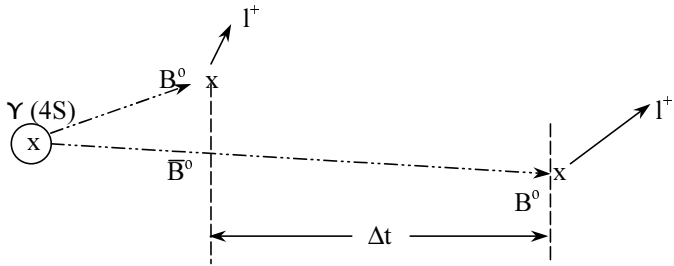,width=0.65\textwidth}
\caption{The flavour tag of $\overline{B}^{o}$ and its decay as 
$B^{o}$ after a time $\Delta t$.}
\label{fig1}
\end{center}
\end{figure}

The case $(l^{+},l^{+})$ associated with the $\overline{B}^{o}\rightarrow 
B^{o}$ transition is shown in Figure \ref{fig1}. 

We consider genuine asymmetries for CP, T and CPT operations, in the sense
that a non-vanishing value is a proof of the violation of the symmetry. Two
cases are of particular interest:

i) {\bf Flavour-to-flavour transitions}

The final configuration denoted by ($l,l$), with flavour definite (for
example, semileptonic) decays detected on both sides of the detector,
corresponds to flavour-to-flavour transitions at the meson level. The
equivalence is shown in Table \ref{tabla1}. The first two processes in the Table are conjugated under CP and also under T.

\bigskip

\begin{table}
\begin{center}
\caption{Flavour-to-flavour transitions}\label{tabla1}\vspace{1ex}
\begin{tabular}[t]{ccc}
\hline
$(X,Y)$ &  & Meson Transition \\ \hline
$(l^{+},l^{+})$ &  & $\overline{B}^{o}\rightarrow B^{o} $ \\ 
  &  & \hspace{7ex} $CP,T$ \\ 
 $(l^{-},l^{-})$ &  & $B^{o}\rightarrow \overline{B}^{o}\hookleftarrow $ \\ 
 $(l^{+},l^{-})$ &  & $\overline{B}^{o}\rightarrow \overline{B}^{o} $
\\ 
  &  &  \hspace{6ex} $CP,CPT$ \\ 
\ $(l^{-},l^{+})$ &  & $B^{o}\rightarrow B^{o}\hookleftarrow $ \\ \hline
\end{tabular}
\end{center}
\end{table}

\bigskip The corresponding Kabir asymmetry \cite{cita06}, to linear order in the CPT violating parameter $\delta $ of the meson mixing, is given by
\begin{equation}
A(l^{+},l^{+})\simeq \frac{4\frac{\text{Re}(\varepsilon)}{1+\left|
\varepsilon\right|^{2}}}{1+4\frac{\text{Re}(\varepsilon)}{1+\left|
\varepsilon\right|^{2}}}\label{2}
\end{equation}
\bigskip
where $\varepsilon $ is the rephasing invariant \cite{cita07} CP-odd, T-odd parameter in the neutral meson mixing. The asymmetry (\ref{2}) does not depend on time. However, in the exact limit $\Delta \Gamma =0$, $\text{Re}(\varepsilon )$ also
vanishes and $A$ will be zero. For the $B_{d}$ system, experimental limits
on $\text{Re}(\varepsilon )$ are of few parts in a thousand \cite{cita08,cita09}.
A second asymmetry arises from the last two processes in Table \ref{tabla1}, related by a CP or a CPT operation.
\begin{equation}
A(l^{+},l^{-})\simeq -2\frac{\text{Re}\left( \frac{\delta }{1-\varepsilon
^{2}}\right) \sinh \frac{\Delta \Gamma \Delta t}{2}-\text{Im}\left( \frac{
\delta }{1-\varepsilon ^{2}}\right) \sin \left( \Delta m\Delta t\right) }{
\cosh \frac{\Delta \Gamma \Delta t}{2}+\cos \left( \Delta m\Delta t\right) }
\label{3}
\end{equation}
\bigskip
which is an odd function of $\Delta t$. This asymmetry also vanishes with $\Delta \Gamma =0$, because then $\text{Im}(\delta )=0$ as well. Present
limits on $\text{Im}(\delta )$ are at the level of few percent \cite{cita08}.

One discovers the weakness of these asymmetries to look for T-, and CPT-
violation in the B$_{d}$-system. The reason is that one needs both the
violation of the symmetry and $\Delta \Gamma \neq 0$.

ii) \ {\bf CP-to-flavour transitions}

Alternative asymmetries can be constructed making use of the CP eigenstates,
which can be identified in this system by means of a CP tag. If the first
decay product, X, is a CP eigenstate produced along the CP-conserving
direction \cite{cita05}, the decay is free of CP violation. If Y is a flavour 
definite channel, then the mesonic transition corresponding to the 
configuration (X,Y) is of the type CP-to-flavour.

In Table \ref{tabla2} we show the mesonic transitions, with their related final
configurations, connected by genuine symmetry transformations to 
$B_{+}\rightarrow B^{o}$.

\begin{table}
\begin{center}
\caption{Transitions connected to $\left( J/\psi K_{S},l^{+}\right) $}\label{tabla2}\vspace{1ex}
\begin{tabular}{ccc}
\hline (X,Y) & Transition & Transformation \\ \hline
$\left( J/\psi K_{S},l^{-}\right) $   & $B_{+}\rightarrow \overline{B}^{o} $
 & CP \\ 
 $\left( l^{-},J/\psi K_{L}\right) $  & $B^{o}\rightarrow B_{+}$ & T \\ 
 $\left( l^{+},J/\psi K_{L}\right) $   & $\overline{B}^{o}\rightarrow B_{+}$ 
& CPT \\ \hline
\end{tabular}
\end{center}
\end{table}

Comparing the intensities of the four processes, we may construct three
genuine asymmetries, namely A(CP), A(T) and A(CPT) \cite{cita10}:

\begin{equation}
A(CP)=-2\frac{\text{Im}(\varepsilon )}{1+\left| \varepsilon \right| ^{2}}%
\sin (\Delta m\Delta t)+\frac{1-\left| \varepsilon \right| ^{2}}{1+\left|
\varepsilon \right| ^{2}}\frac{2\text{Re}(\delta )}{1+\left| \varepsilon
\right| ^{2}}\sin ^{2}\left( \frac{\Delta m\Delta t}{2}\right)  \label{4}
\end{equation}

\bigskip

The CP-odd asymmetry, contains both T-violating and CPT-violating
contributions, which are, respectively, odd and even functions of $\Delta t$%
. This asymmetry corresponds to the ''gold plate'' decay \cite{cita11} and has been
measured recently \cite{cita01}. The result is interpreted in terms of the standard
model $\sin (2\beta )$ with neither CPT-violation nor $\Delta \Gamma $ \cite{cita12}.
One finds \cite{cita07}
\begin{equation}
-\frac{2\text{Im}(\varepsilon )}{1+\left| \varepsilon \right| ^{2}}=\sin
(2\beta )  \label{5}
\end{equation}

The two T- and CPT-violating terms in Eq. (\ref{4}) can be separated out by
constructing other asymmetries

\begin{equation}
A(T)=-2\frac{\text{Im}(\varepsilon )}{1+\left| \varepsilon \right| ^{2}}\sin
(\Delta m\Delta t)\left[ 1-\frac{1-\left| \varepsilon \right| ^{2}}{1+\left|
\varepsilon \right| ^{2}}\frac{2\text{Re}(\delta )}{1+\left| \varepsilon
\right| ^{2}}\sin ^{2}\left( \frac{\Delta m\Delta t}{2}\right) \right]
\label{6}
\end{equation}

the T-asymmetry needs $\varepsilon \neq 0$ and turns out to be purely odd in 
$\Delta t$ in the limit we are considering.

\begin{equation}
A(CPT)=\frac{1-\left| \varepsilon \right| ^{2}}{1+\left| \varepsilon \right|
^{2}}\frac{2\text{Re}(\delta )}{1+\left| \varepsilon \right| ^{2}}\frac{\sin
^{2}\left( \frac{\Delta m\Delta t}{2}\right) }{1-2\frac{\text{Im}%
(\varepsilon )}{1+\left| \varepsilon \right| ^{2}}\sin (\Delta m\Delta t)}
\label{7}
\end{equation}

is the CPT asymmetry. It needs $\delta \neq 0$ and includes both even and
odd time dependences.

The expressions (\ref{4}), (\ref{6}) and (\ref{7}) correspond to the limit $\Delta \Gamma =0$,
but, being genuine observables, a possible absorptive part could not induce
by itself a non-vanishing asymmetry.\bigskip \bigskip

\section{$\Delta \Gamma $ and Non-genuine Asymmetry}\label{section3}

The construction of the quantities described above requires to tag both B$%
_{+}$ and B$_{-}$ states, and thus the reconstruction of both $B\rightarrow
J/\psi K_{S}$ \ and $B\rightarrow J/\psi K_{L}$ decays. One can consider
non-genuine asymmetries from $B\rightarrow J/\psi K_{S}$ only: they involve
\cite{cita10} the discrete transformation, denoted by $\Delta t$, consisting 
in the exchange in the order of appearance of decay products X and Y, 
which cannot be associated with any fundamental symmetry.

Table \ref{tabla3} shows the different transitions we may study from such final states.
Besides the genuine CP asymmetry, there are two new quantities that can be
constructed from the comparison between $\left( J/\psi K_{S},l^{+}\right) $
and the processes in the Table \ref{tabla3}.

\begin{table}
\begin{center}
\caption{Final configurations with only $J/\psi K_{S}$}\label{tabla3}\vspace{1ex}
\begin{tabular}{ccc}
\hline
(X,Y) & Transition & Transformation \\ \hline
$\left( J/\psi K_{S},l^{+}\right) $ & $B_{+}\rightarrow B^{o}$ & $CP$ \\ 
\ $\left( l^{+},J/\psi K_{S}\right) $ & $\overline{B}^{o}\rightarrow B_{-}$
& $\Delta t$ \\ 
\ $\left( l^{-},J/\psi K_{S}\right) $ & $B^{o}\rightarrow B_{-}$ & $CP\Delta
t$ \\ \hline
\end{tabular}
\end{center}
\end{table}

In the exact limit $\Delta \Gamma =0$, $\Delta t$ and T operations, although
different (compare the second lines of Tables \ref{tabla2} and \ref{tabla3}), are found to become
equivalent, so that the temporal asymmetries satisfy $A(\Delta t)=A(T)$ and 
$A(CP\Delta t)=A(CPT)$.

The asymmetries $A(\Delta t)$ and $A(CP\Delta t)$ are non-genuine, so that
the presence of $\Delta \Gamma \neq 0$ may induce non-vanishing values for
them, even in the absence of true T or CPT violation. These effects can be
calculated and are thus controllable. This reasoning leads to an interesting
suggestion: {\bf there are linear terms in }$\Delta \Gamma ${\bf \ inducing
a non-vanishing asymmetry }${\bf A(CP\Delta t)}$. This last asymmetry is
particularly clean, under the reasonable assumption that $A(CPT)=0$.
Explicit calculations \cite{cita10} show that, even in the limit of perfect symmetry,
i.e., $\varepsilon =0$ besides $\delta =0$, one finds a non-vanishing $%
(CP\Delta t)$-asymmetry, given by

\begin{equation}
A(l^{-},J/\psi K_{S})=\frac{\Delta \Gamma \Delta t}{2}\text{; \ \ }%
\varepsilon =\delta =0  \label{8}
\end{equation}

The simple result (\ref{8}) is modified under the realistic $\varepsilon \neq 0$
situation. One has, if only $\delta =0$, the result

\begin{equation}
A(l^{-},J/\psi K_{S})=\frac{1}{1-\frac{2\text{Im}(\varepsilon )}{1+\left|
\varepsilon \right| ^{2}}\sin (\Delta m\Delta t)}  \label{9}
\end{equation}

\[
\left\{ \frac{\Delta \Gamma \Delta t}{2}\frac{1-\left| \varepsilon \right|
^{2}}{1+\left| \varepsilon \right| ^{2}}+\frac{4\text{Re}(\varepsilon )}{%
1+\left| \varepsilon \right| ^{2}}\sin ^{2}\left( \frac{\Delta m\Delta t}{2}%
\right) -\frac{2\text{Im}(\varepsilon )}{1+\left| \varepsilon \right| ^{2}}%
\frac{2\text{Re}(\varepsilon )}{1+\left| \varepsilon \right| ^{2}}\sin
(\Delta m\Delta t)\right\} 
\]

\bigskip

The three terms of Eq. (\ref{9}) contain different $\Delta t$-dependences, so that
a good time resolution will allow the determination of the parameters.
Taking into account that $\text{Re}(\varepsilon )=x\Delta \Gamma $, all of
the three terms are linear in $\Delta \Gamma $. We conclude that the
comparison between the channels $\left( l^{-},J/\psi K_{S}\right) $ and $%
\left( J/\psi K_{S},l^{+}\right) $ is a good method to obtain information on 
$\Delta \Gamma $, due to the absence of any non-vanishing difference when $%
\Delta \Gamma =0$.
\bigskip
\section{$B_{s}$ Mixing}\label{section4}
\bigskip

\begin{figure}[h]
\begin{center}
\epsfig{figure=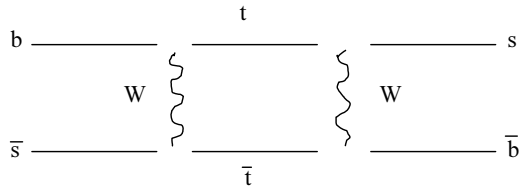,width=0.5\textwidth}
\caption{Box Diagram responsible of the neutral meson mixing}
\label{fig2}
\end{center}
\end{figure}

$B^{o}$ and $\bar{B}^{o}$ are not mass eigenstates, so that their
oscillation frequency is governed by their mass-difference. The measurement
by the UA1 collaboration \cite{cita13} of a large value of $\Delta M_{d}$ was
historically the first indication of the heavy top quark mass. \ This is so
because of non-decoupling effects of the heavy-mass exchange in the Box
Diagram. For $B_{s}-mixing$ , it is shown in Figure \ref{fig2}.
\bigskip

\bigskip

To avoid many hadronic uncertainties, it is interesting to consider the
ratio between $\Delta M_{s}$ and $\Delta M_{d}$, given by \cite{cita14}

\begin{equation}
\frac{\Delta M_{s}}{\Delta M_{d}}=\left| \frac{V_{ts}}{V_{td}}\right| ^{2}%
\frac{M_{Bd}}{M_{Bs}}\xi ^{2}  \label{10}
\end{equation}

where $\xi ^{2}\equiv \frac{f_{B_{s}}^{2}B_{B_{s}}}{f_{B_{d}}^{2}B_{B_{d}}}$
\ is the flavour-SU(3) breaking parameter in terms of the meson decay
constants and the bag factors. $B_{B_{q}}=1$ \ if one uses a vacuum
saturation of the hadronic matrix element. The great advantage of Eq. (\ref{10})
is that, in the ratio, different systematics in the evaluation of the matrix
element tends to cancel out. However, unlike $\Delta M_{d}=0.479(12)$ $%
ps^{-1}$, which is measured with a good precision \cite{cita15}, the determination of 
$\Delta M_{s}$ is an experimental challenge due to the rapid oscillation of
the $B_{s}-system$. At present \cite{cita15}, $\Delta M_{s}>13.1ps^{-1}$, with 95\%\
C.L., but this bound already provides a strong constraint on $\left|
V_{td}\right| $. The use of QCD spectral sum rules leads to \cite{cita14}.
\begin{equation}
\xi \simeq 1.18\pm 0.03\rightarrow \Delta M_{s}\simeq 18.6(2.1)\text{ps}^{-1}
\label{11}
\end{equation}

in agreement with the present experimental lower bound and within the reach
of the proposed experiments.

Ali and London \cite{cita16} have examined the situation for SUSY theories with
minimal flavour violation. In this class of models, the SUSY contributions
to $\Delta M_{d}$ and $\Delta M_{s}$ can both be described by a single
common parameter $f$.

\begin{eqnarray}
\Delta M_{d} &=&\Delta M_{d}(SM)[1+f]  \label{12} \\
\Delta M_{s} &=&\Delta M_{s}(SM)[1+f]
\end{eqnarray}

The parameter $f$ is positive definite, so that the SUSY contributions add
constructively to the SM contributions in the entire allowed supersymmetric
parameter space. The size of $f$ depends, in general, on the parameters of
the SUSY model. They conclude that, if $M_{s}$ is measured to be near its
lower limit, SUSY with large $f$ is disfavoured.

With respect to the values of the CP phases $\alpha $, $\beta $ and $\gamma $
of the unitarity triangle, the key observation is that a measurement of $%
\beta $ will not distinguish among the various values of $f$, i.e., $\beta $
is rather independent of $f$. If one wants to distinguish among the various
SUSY models, it will be necessary to measure $\gamma $ and/or $\alpha $
independently.

Contrary to these SUSY models, in which the ratio $\Delta M_{s}/\Delta M_{d}$
remains that of the Standard Model, Left-Right-Symmetric Models with
Spontaneous CP Violation modify $\Delta M_{s}$ and $\Delta M_{d}$ with
different phases relative to the SM contribution. One has \cite{cita17}

\begin{equation}
\frac{\Delta M_{s}}{\Delta M_{d}}=\frac{\Delta M_{s}(SM)}{\Delta M_{d}(SM)}%
\left| \frac{1+\kappa e^{i\sigma _{s}}}{1+\kappa e^{i\sigma _{d}}}\right| 
\label{13}
\end{equation}

As a consequence, the ratio is modified with respect to the SM. In Ref.
\cite{cita18}, an analysis of the joint constraints imposed by $\Delta M_{K}$ 
and $\Delta M_{B}$ is performed, with the conclusion that the Left-Right Model
favours opposite signs of $\epsilon _{K}$ and $\sin (2\beta)$ and it would
be disfavoured for $\sin (2\beta)>0.1$. A test of the $B_{s}-mixing$ would
be crucial in this context.

\section{Two Decays: $B_{s}\rightarrow J/\protect\psi \protect\phi $, $
B_{d}\rightarrow \protect\phi K_{S}$}\label{section5}

\begin{figure}[ht]
\begin{center}
\epsfig{figure=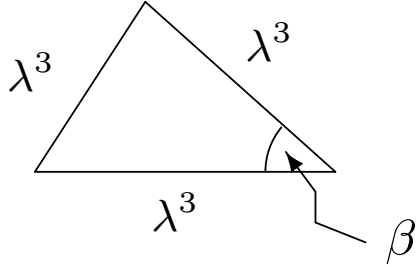,width=0.4\textwidth}
\caption{The ($bd$) unitarity triangle.}
\label{fig3}
\end{center}
\end{figure}

The general argument of considering CP-Violation in $B_{s}-mixing$ as a
prime candidate for New Physics is well defined. In $B_{d}\rightarrow J/\psi
K_{S}$, the SM amplitudes of mixing (dominated by the virtual top quark) $%
V_{bt}V_{td}^{\ast }\sim \lambda ^{3}$ and decay $V_{bc}V_{cd}^{\ast }\sim
\lambda ^{3}$ define a relative phase $\beta $ of the order of one, because
the corresponding unitarity triangle satisfies the scales of figure \ref{fig3}.

\begin{figure}[hb]
\begin{center}
\epsfig{figure=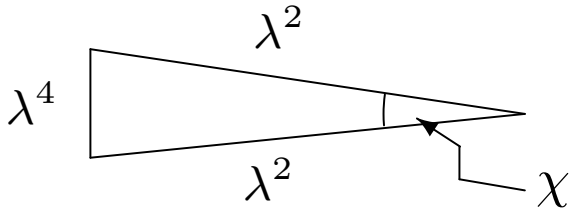,width=0.55\textwidth}
\caption{The ($bs$) unitarity triangle.}
\label{fig4}
\end{center}
\end{figure}

Contrary to this standard $\beta $, the SM amplitudes for $B_{s}\rightarrow
J/\psi \phi $ satisfy that $V_{bt}V_{ts}^{\ast }\sim \lambda ^{2}$, $
V_{bc}V_{cs}^{\ast }\sim \lambda ^{2}$, so that the corresponding unitarity
triangle is shown in figure \ref{fig4} and the relative phase $\chi $ is tiny, of the order of $\lambda ^{2}$. It
does not take much New Physics to change the tiny standard $\chi $!

In hadronic machines, substantial number of $B_{s}\rightarrow J/\psi \phi $
events are expected. The decay process is described by three Lorentz
invariant terms, two CP-even terms and one CP-odd term. The joint angular
distibution with $J/\psi \rightarrow l^{+}l^{-}$ and $\phi \rightarrow
K^{+}K^{-}$ has been described in Ref. \cite{cita19}, with the aim of 
separating out the definite CP eigenstates and thus recovering a CP asymmetry 
free of cancellations.

Following the argument of the Introduction, candidates for New Physics are
processes which, in the SM, are described by either mixing or penguin
amplitudes in the decay. In Left-Right models, the gluonic penguin
contribution to $b\rightarrow s\overline{s}s$ transition is enhanced by $%
m_{t}/m_{b}$ due to the presence of right-handed currents. This may overcome
the suppression due to small left-right mixing angle. Two new phases \cite{cita20} in
the $B\rightarrow \phi K_{S}$ decay amplitude may therefore modify the time
dependent CP asymmetry in this decay mode by O(1). This scenario implies
also large CP asymmetry in the decay $B_{s}\rightarrow \phi \phi $ which can
be tested in hadronic machines.

\section{Conclusions}\label{section6}

The prospects for an experimental study of the Flavour Problem in the next
coming years are much interesting, from CP-violating observables in
B-factories and hadronic machines. The (bd) Unitarity Triangle will be
tested, with separate determinations of the CP-phases ($\alpha $, $\beta $, 
$\gamma $), after the present establishment of CP-Violation in $B_{d}$
-physics.

New phenomena are probably around the corner. Discoveries like T-violation
in B-physics, without any contamination of absorptive parts, and sensitive
limits (or spectacular surprises) for CPT-violation are expected. Temporal
asymmetries in B-decays are a good method to search for linear terms in $
\Delta \Gamma /\Gamma $. The intensities $I(l^{-}$, $J/\psi K_{S})$ and $
I(J/\psi K_{S}$, $l^{+})$ are predicted to be equal under CPT invariance and 
$\Delta \Gamma =0$. Linear terms in $\Delta \Gamma $ induce a non-vanishing
asymmetry for this CP $\Delta t$ transformation.

$B_{s}-mixing$ is considered to be of prime importance for the search of new
physics, particularly in its CP-violating component. Extended models modify
the tiny phase between the top and charm sides of the standard (bs)
unitarity triangle. The non-decoupling effects of new physics can be put
under control by the cancellation of hadronic matrix element uncertainties
in the ratio $\Delta M_{s}/\Delta M_{d}$. The interest in a detailed
analysis of $B_{s}\rightarrow J/\psi \phi $ is apparent in this context.

New physics in penguin-mediated decays, like $B_{d}\rightarrow \phi K_{S}$
and $B_{s}\rightarrow \phi \phi $, is also expected, with information
complementary to that of mixing.

All in all, we can expect a beautiful future in front of us!

\bigskip

\section*{Acknowledgements}\label{section7}

I would like to thak E. Alvarez, P. Ball, D. Binosi, F.Botella, J. Matias, M.Nebot and J. Papavassiliou for interesting discussions and to the Organizers of
''Beauty 2002'' for their invitation to such a beautiful event in Santiago.
This work has been supported by the Grant AEN-99/0692 of the Spanish
Ministry of Science and Technology.

\end{document}